\documentclass[aps,prstab,twocolumn]{revtex4-1}
\usepackage{graphicx}
\usepackage{dcolumn}
\usepackage{bm}
\usepackage{color}
\usepackage{amsmath}
\usepackage{gensymb}
\usepackage{natbib}

\begin{document}
\title{Interface enhanced spin-orbit torques and current-induced magnetization switching of Pd/Co/AlO$_x$ layers}

\author{Abhijit Ghosh}
\author{Kevin Garello}
\author{Can Onur Avci}
\author{Mihai Gabureac}
\author{Pietro Gambardella}
\affiliation{Department of Materials, ETH Z\"{u}rich, H\"{o}nggerbergring 64, CH-8093 Z\"{u}rich, Switzerland}

\date{\today}

\begin{abstract}
Magnetic heterostructures that combine large spin-orbit torque efficiency, perpendicular magnetic anisotropy, and low resistivity are key to develop electrically-controlled memory and logic devices. Here we report on vector measurements of the current-induced spin orbit torques and magnetization switching in perpendicularly magnetized Pd/Co/AlO$_x$ layers as a function of Pd thickness. We find sizeable damping-like (DL) and field-like (FL) torques, of the order of 1~mT per $10^7$~A/cm$^2$, which have different thickness and magnetization angle dependence. The analysis of the DL torque efficiency per unit current density and electric field using drift-diffusion theory leads to an effective spin Hall angle and spin diffusion length of Pd larger than 0.03 and 7~nm, respectively. The FL SOT includes a significant interface contribution, is larger than estimated using drift-diffusion parameters, and is further strongly enhanced upon rotation of the magnetization from the out-of-plane to the in-plane direction. Finally, taking advantage of the large spin-orbit torques in this system, we demonstrate bipolar magnetization switching of Pd/Co/AlO$_x$ layers with similar current density as used for Pt/Co layers with comparable perpendicular magnetic anisotropy.
\end{abstract}
\maketitle

\section{INTRODUCTION}
Spin-orbit torques (SOT) generated by current injection in normal metal/ferromagnet (NM/FM) bilayers have attracted considerable attention as a method to induce magnetization switching of thin FM films and magnetic tunnel junction devices~\cite{MironN2011,AvciAPL2012,LiuS2012,CubukcuAPL2014,GarelloAPL2014,Beach2014APL,zhang2015apl}. Although the bulk or interface origin of the spin current giving rise to the damping-like (DL) and field-like (FL) SOT components is a matter of debate~\cite{HaneyPRB2013,Freimuth2014PRB,FreimuthPRB2015a,KellyPRL2016}, it has been established that the strongest SOT are found in NM with large spin-orbit coupling, and hence large spin Hall effect (SHE) and small spin diffusion length ($\lambda$), such as the $5d$-metals Pt, W, and Ta~\cite{LiuPRL2011,PaiAPL2012,GarelloNN2013,KimNM2013,AvciPRB2014,FanNC2014,torrejon2014nc,Avci2015APL}. For this reason, the $4d$-metals have been largely cast aside in the quest for large SOT, even though several studies of the SHE in these materials show sizeable spin Hall angles, of the order of $\gamma_{SH} \sim 0.01$~\cite{Otani-PRB-2011,KondouAPEX2012,tang12apex,Vlaminck2013prb,Silva2013JAP}.

Pd-based heterostructures constitute an apparent exception to this trend: recent studies of [Pd/Co]$_\mathrm{N}$ multilayers~\cite{Jamali2013PRL} and Pd/FePd/MgO trilayers~\cite{Lee2014SR} evidenced DL and FL effective fields of the same order of magnitude or larger than those found in Pt- and Ta-based structures, reaching up to more than 10~mT for an injected current density of $j=10^7$~A/cm$^2$. The effective spin Hall angle required to model the SOT according to spin diffusion theory is $\gamma_{SH} >1$ in [Pd/Co]$_\mathrm{N}$ (Ref.~\onlinecite{Jamali2013PRL}) and $\gamma_{SH} \approx 0.15 $ in Pd/FePd/MgO~\cite{Lee2014SR}, which is substantially larger compared to previous estimates of $\gamma_{SH}$ in Pd~\cite{Otani-PRB-2011,KondouAPEX2012,tang12apex,Vlaminck2013prb,Silva2013JAP}. This comparison calls for further investigations of the SOT due to Pd as well as of the possibility of using Pd to induce magnetization switching, which has not been reported to date.

In this work we present a study of current-induced SOT and magnetization switching in Pd/Co/AlO$_x$ trilayers with varying Pd thickness. We focus on the simple trilayer structure to avoid the complexity due to multiple interfaces and current paths in [Pd/Co]$_\mathrm{N}$ multilayers, which complicate the analysis and interpretation of the experimental data~\cite{Jamali2013PRL}. Our study is based on the harmonic Hall voltage method to measure SOT~\cite{GarelloNN2013,KimNM2013,AvciPRB2014,PiAPL2010,HayashiPRB2014} and includes a detailed magnetotransport characterization as well as the analysis of magnetothermal contributions to the Hall resistance~\cite{GarelloNN2013,AvciPRB2014b}. We report strong DL and FL SOT amplitudes and efficient magnetization switching for this system, combined with large perpendicular magnetic anisotropy (PMA) and relatively low resistivity. We further observe a distinct thickness dependence of the DL and FL torques and discuss different models to account for the SOT efficiency. The main results are summarized at the end of this paper.

\section{EXPERIMENT}
Our samples are Ta(0.8nm)/Pd($t_{Pd}$)/Co(0.6nm)/ AlO$_x$(1.6nm) layers grown on thermally oxidized Si substrates by dc magnetron sputtering in an Ar pressure of $2\times 10^{-3}$ Torr. The Al cap layer was oxidized for 37~s under a partial O$_2$ pressure of $7\times 10^{-3}$~Torr in order to increase the PMA of Co/Pd. The Ta seed layer was used to improve the uniformity of Pd; because of its high resistivity and reduced thickness, such a layer plays a negligible role in the generation of SOT, as we confirmed by measuring Pd/Co/AlO$_x$ structures grown with a Ti seed. The rms roughness measured by atomic force microscopy was found to vary between 0.15~nm ($t_{Pd}=1.5$~nm) and 0.13~nm ($t_{Pd}=8$~nm). The as-grown layers were patterned using UV-lithography and dry etching into Hall bars of width $w=10$~$\mu$m and length $L=50$~$\mu$m, as shown in Fig.~\ref{fig1}a. For the magnetotransport characterization, the samples were mounted in the gap of an electromagnet producing a field $B_{ext}$ and allowing for rotating the current direction relative to the magnetization (\textbf{m}). The experiments were performed at room temperature using an ac current $I=I_{0}\sin(\omega t)$ with amplitude $I_0$ and frequency $\omega/2\pi=10$~Hz. $I_0$ ranged between 3 and 14~mA in samples with $t_{Pd}=1.5$ and 9~nm, respectively, and was adjusted in order to keep the current density flowing through Pd between $j_{Pd}=1\times 10^7$ and $1.5\times 10^7$~A/cm$^2$.
The harmonic Hall voltage measurements were performed by taking the Fourier transform of the ac Hall resistance recorded for 10~s at each value of the applied field and current. The first harmonic Hall resistance is analogous to a dc Hall measurement and is given by 
\begin{equation}\label{eq1stharmonic}
R^{\omega}_{H}=R_{AHE}\cos\theta + R_{PHE}\sin^2\theta\sin (2\phi),
\end{equation}
where $R_{AHE}$ and $R_{PHE}$ are the anomalous and planar Hall coefficients, and $\theta$ and $\phi$ represent the polar and azimuthal angles of the magnetization, respectively. Current-induced effects were characterized by measuring the second harmonic Hall resistance, $R^{2\omega}_{H}$, which arises due to the mixing of the ac current with the Hall effect modulated by the oscillations of the magnetization induced by the DL and FL torques~\cite{GarelloNN2013,KimNM2013,AvciPRB2014,PiAPL2010,HayashiPRB2014}.
This term accounts for the SOT effective fields, which are proportional to $I$, as well as for magnetothermal effects that are proportional to $I^2$~\cite{GarelloNN2013,AvciPRB2014,AvciPRB2014b}. Following Ref.~\onlinecite{AvciPRB2014b}, we have
\begin{multline}
\label{eq2}
R^{2\omega}_{H}=[R_{AHE}-2R_{PHE}\cos\theta\sin(2\phi)]\frac{d\cos\theta}{dB_{ext}} \frac{B_{\theta}^{I}}{\sin(\theta_B-\theta)} \\
+R_{PHE}\sin^{2}\theta\frac{2\cos(2\phi)}{B_{ext}\sin\theta_B}B_{\phi}^{I}+ R^{2\omega}_{\nabla T } \, ,
\end{multline}
where $B_{\theta}^{I}$ and $B_{\phi}^{I}$ are the polar and azimuthal components of the current-induced effective fields, respectively, including both SOT and Oersted field, and $R^{2\omega}_{\nabla T} \propto (\nabla T \times \mathbf{m})\cdot \mathbf{y}$ is the thermal Hall resistance associated to the anomalous Nernst effect. The latter is caused by the unintentional out-of-plane and in-plane temperature gradients $\nabla T $ produced by Joule heating and asymmetric heat dissipation in the trilayer~\cite{AvciPRB2014b}. To separate the SOT contributions from $R^{2\omega}_{\nabla T}$, we exploit the different behavior of $B_{\theta,\phi}^{I}$ and $R^{2\omega}_{\nabla T}$ as a function of $B_{ext}$, as explained in detail in Refs.~\onlinecite{AvciPRB2014b} and \onlinecite{SI}.
%
Finally, in order to explicit the relationship between $B_{\theta}^{I}$ and $B_{\phi}^{I}$ in Eq.~\ref{eq2} and the DL and FL SOT effective fields, we express the latter in spherical coordinates~\cite{GarelloNN2013}:
\begin{eqnarray}
\label{eq3-4}
\mathbf{B}^{DL} = B^{DL}_\theta \cos\phi \, \mathbf{e}_{\theta} - B^{DL}_\phi \cos\theta\sin\phi \, \mathbf{e}_{\phi}, \\
\mathbf{B}^{FL} = -B^{FL}_\theta\cos\theta\sin\phi \, \mathbf{e}_{\theta} -B^{FL}_\phi \cos\phi \, \mathbf{e}_{\phi},
\end{eqnarray}
where the coefficients $B^{DL}_{\theta}$, $B^{FL}_{\theta}$, $B^{DL}_{\phi}$, and $B^{FL}_{\phi}$ represent the polar and azimuthal amplitudes of the DL and FL SOT. In the "isotropic torque" limit usually considered in the SOT literature, Eqs.~\ref{eq3-4},~4 are obtained by projecting $\mathbf{B}^{DL} = B^{DL}_0 (\mathbf{m}\times \mathbf{y})$ and $\mathbf{B}^{FL} = B^{FL}_0 [\mathbf{m}\times(\mathbf{m}\times \mathbf{y})]$ onto the unit vectors $\mathbf{e}_{\theta}$ and $\mathbf{e}_{\phi}$ and by assuming $B^{DL}_{\theta,\phi} = B^{DL}_{0}$ and $B^{FL}_{\theta,\phi} = B^{FL}_{0}$. General models and SOT vector measurements, however, have shown that $B^{DL}_{\theta,\phi}$ and $B^{FL}_{\theta,\phi}$ can be functions of the magnetization orientation~\cite{GarelloNN2013,KJLee14SR,KJLee15PRB}. When this occurs, as in the present case, the four unknown coefficients in Eqs.~\ref{eq3-4},~4 must be determined by four independent measurements of $R_H^{2\omega}$ as a function of $B_{ext}$, applied in-plane along $\phi=0 \degree, \pm 45 \degree$ and $90 \degree$.
All the SOT values reported in this work have been normalized to $j_{Pd}=1\times 10^7$~A/cm$^2$ for comparison with literature data.

\section{RESULTS}
\subsection{Resistivity and magnetic anisotropy}
\begin{figure*}
	\centering
    \includegraphics[width=12 cm]{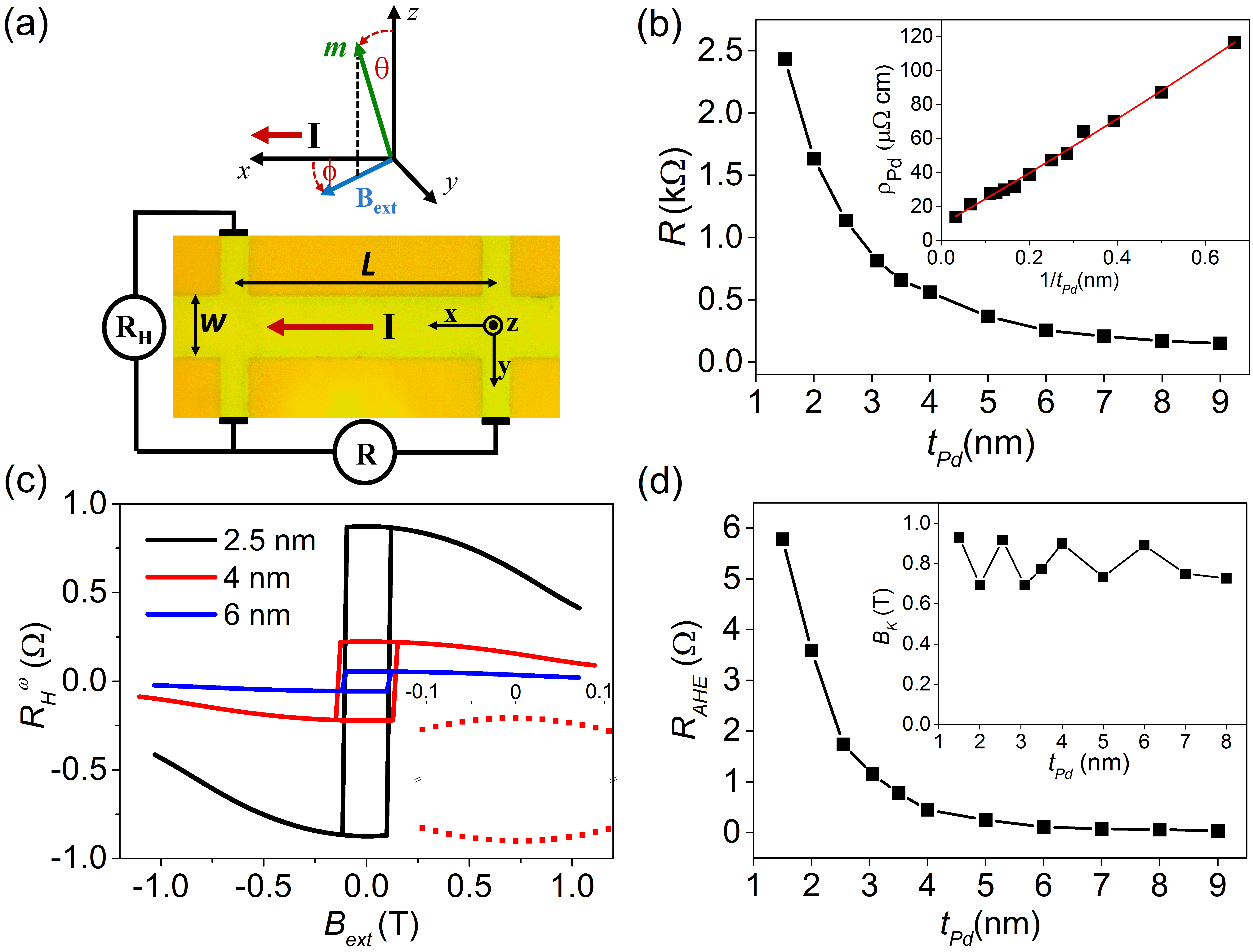}
    \caption{(a) Schematic of the experimental geometry and scanning electron micrograph of a Pd/Co/AlO$_x$ Hall bar. (b) Resistance as a function of Pd thickness. Inset: Pd resistivity as a function of inverse thickness; the solid line is a linear fit. (c) First harmonic Hall resistance $R^{\omega}_{H}$ of three representative samples as a function of external field applied at $\theta_B = 84^{\circ}$ and $\phi = 0^{\circ}$. Inset: detail of the low field region of the 4~nm sample. (d) $R_{AHE}$ as a function of Pd thickness. Inset: Anisotropy field $B_K$.}\label{fig1}
\end{figure*}

Figure~\ref{fig1}(b) shows the resistance of the Pd/Co/AlO$_x$ layers as a function of Pd thickness. Using a simple parallel resistor model, assuming that the conductivity of Co remains constant across the series, we find that the resistivity of Pd varies approximately as $\rho_{Pd}\propto 1/t_{Pd}$ [inset of Fig.~\ref{fig1}(b)]. This is in line with previous investigations of the resistivity of Pd thin films~\cite{ShivaprasadPL80}, which behaves similarly to Pt~\cite{Fuchs-PRB_1980}, and is significantly smaller compared to $\beta$-W and $\beta$-Ta. The Hall resistance $R^{\omega}_{H}$ is shown in Fig.~\ref{fig1}(c) as a function of $B_{ext}$ applied at an angle $\theta_B = 84^{\circ}$ relative to the sample normal. As customary in SOT investigations of PMA materials, the external field was slightly tilted off-plane in order to prevent the formation of magnetic domains. Accordingly, we observe that the magnetization loops are reversible as long as \textbf{m} does not switch direction, as expected for coherent rotation of the magnetization, and that $R^{\omega}_{H}$ decreases when increasing $B_{ext}$ as \textbf{m} tilts away from $\mathbf{z}$. When $\phi = 0^{\circ}, 90^{\circ}$, $R^{\omega}_{H}$ is proportional to $\mathbf{m}\cdot \mathbf{z}$, which allows us to derive the angle $\theta=\cos^{-1}(R_H^{\omega}/R_{AHE})$ as a function of applied field. We observe also that the maximum amplitude of $R^{\omega}_{H}$ decreases with increasing Pd thickness owing to the decrease of the AHE due to current shunting through Pd [Fig.~\ref{fig1}(d)]. Our data indicate that both the AHE and PHE are roughly proportional to $t_{Pd}^{-3}$ rather than to $\rho_{Pd}/t_{Pd} \propto t_{Pd}^{-2}$, as would be expected for a parallel resistor model of the Co and Pd layers. This suggests that the AHE and the magnetoresistance are enhanced at the Pd/Co interface relative to bulk Co, in agreement with studies of the AHE in [Co/Pd]$_\mathrm{N}$ multilayers~\cite{guoPRB12}.

The inset of Fig.~\ref{fig1}(d) shows the effective magnetic anisotropy field calculated by using the macrospin approximation: $B_K = B_{ext} (\sin\theta_{B}/\sin\theta -\cos\theta_{B}/\cos\theta)$. We find that $B_K$ varies between 0.7 and 0.9~T, with no systematic variation as a function of Pd thickness and independently of the in-plane direction of the magnetization. The corresponding uniaxial magnetic anisotropy energy is thus $K_u = B_K M_s/2 + \mu_0 M_s^2/2 = (1.5 \pm 0.1) \times 10^6$~J/m$^3$, where $M_s = 1.27 \times 10^6$~A/m is the saturation magnetization measured by SQUID. Interestingly, $K_u$ for Pd/Co/AlO$_x$ is comparable or larger relative to [Pd/Co]$_\mathrm{N}$ multilayers~\cite{Carcia85apl,Fullerton09APL,Ando10apl}, which we attribute to the top oxide layer favoring strong PMA through the formation of Co-O bonds~\cite{yang11prb,rau14science}.

\subsection{Spin-orbit torques}
\begin{figure*}[t]
    \centering
    \includegraphics[width=12cm]{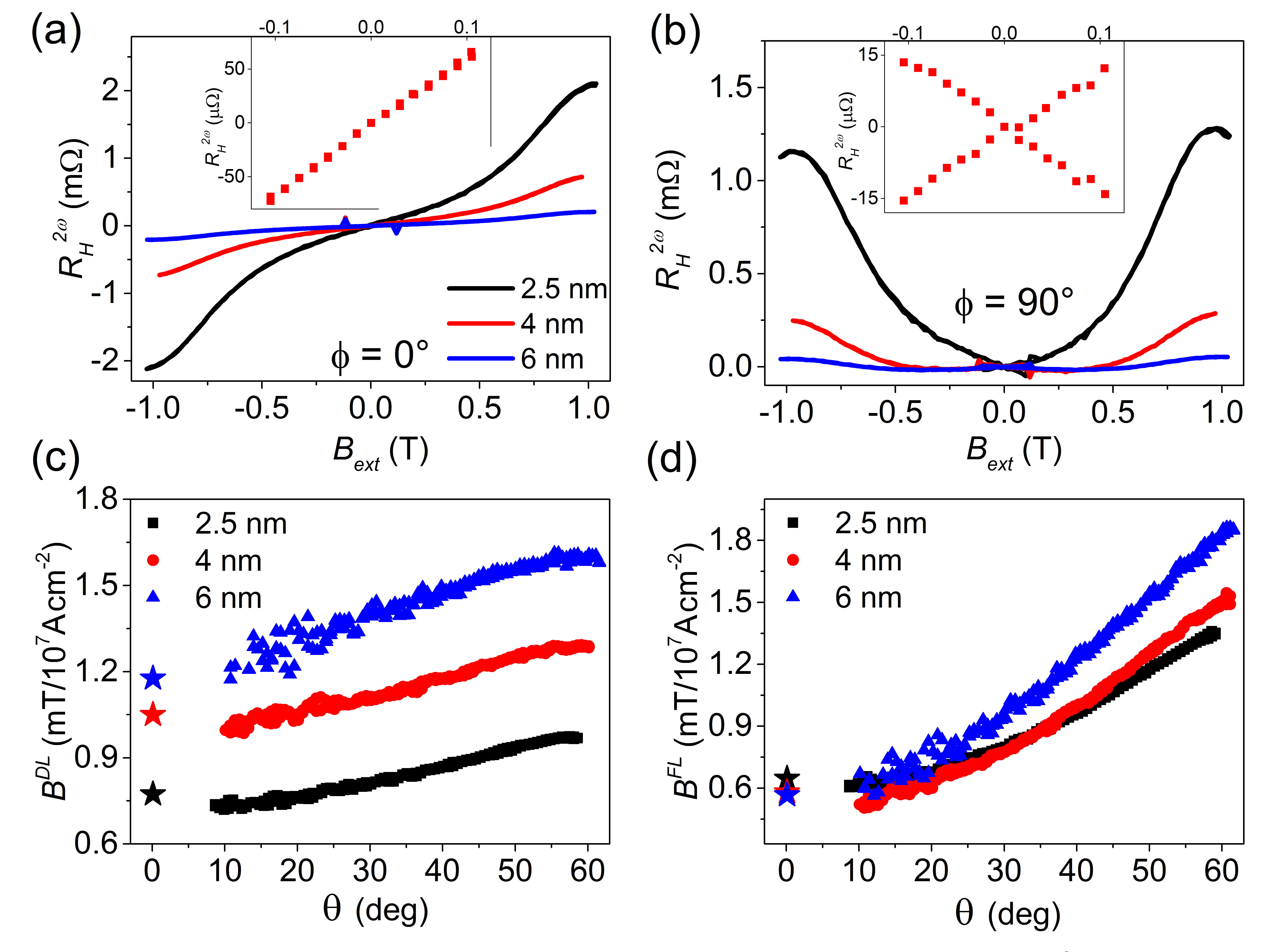}
    \caption{Second harmonic Hall resistance $R^{2\omega}_{H}$ as a function of applied field at $\phi = 0\degree$ (a) and $\phi = 90\degree$ (b). The insets show $R^{2\omega}_{H}$ in the small angle limit ($\theta \leq 7 \degree$).
    (c) $B^{DL}$ and (d) $B^{FL}$ extracted from the data in (a) and (b), respectively, as a function of the polar magnetization angle $\theta$. $B^{DL}$ and $B^{FL}$ obtained using the small angle approximation are indicated by a star.}\label{fig2}
\end{figure*}

The SOT measurements were performed by analyzing the second harmonic Hall resistance that arises due to the mixing of the ac current with the ac oscillations of the magnetization induced by the DL and FL torques (Eq.~\ref{eq2}). Because the DL torque is larger when $\mathbf{m}$ is oriented in the $xz$ plane, whereas the FL torque tends to align $\mathbf{m}$ towards $y$, measurements taken at $\phi = 0\degree$ ($\phi = 90\degree$) reflect mainly the strength of the DL (FL) effective fields. Figures~\ref{fig2}(a) and (b) show $R^{2\omega}_{H}$ measured at $\phi = 0\degree$ and $\phi = 90\degree$ after the subtraction of $R^{2\omega}_{\nabla T}$~\cite{SI}. These curves are, respectively, odd and even with respect to magnetization reversal, reflecting the different symmetry of $B^{DL}$ and $B^{FL}$~\cite{GarelloNN2013}. Similarly to $R^{\omega}_{H}$, we observe that the amplitude of $R^{2\omega}_{H}$ decreases with increasing Pd thickness, which limits the signal-to-noise ratio and, consequently, the range of our SOT measurements to $t_{Pd} \leq 8$~nm.
Figures~\ref{fig2}(c) and (d) show the effective fields $B^{DL}_{\theta}$ and $B^{FL}_{\theta}$ as a function of $\theta$, obtained by $R_H^{2\omega}$ using Eqs.~\ref{eq2}-4. We find that the DL and FL fields have the same sign as measured for Pt/Co/AlO$_x$~\cite{MironN2011,GarelloNN2013,MironNM2010} and increase as the magnetization rotates towards the plane of the layers. Compatibly with the uniaxial symmetry of the system, the angular dependence of the DL and FL fields can be modelled by a Fourier series expansion of the type $B^{DL,FL}_\theta = B^{DL,FL}_0 + B^{DL,FL}_2 \sin^2\theta + B^{DL,FL}_4 \sin^4\theta + ...$, whereas the azimuthal components are only weakly angle dependent and are approximated by $B^{DL,FL}_\phi \approx B^{DL,FL}_0$~\cite{GarelloNN2013}. The values of the zeroth, second, and fourth order $B^{DL,FL}_\theta$ coefficients are obtained by fitting the data in Fig.~\ref{fig2}(c) and (d) according to this expansion. Additionally, we present data obtained using the small angle approximation~\cite{HayashiPRB2014}, which yields accurate values for $B^{DL,FL}_0$ when $\theta \approx 0\degree$ [star symbols in Fig.~\ref{fig2}(c) and (d)]. This approximation is valid as long as $R_H^{2\omega}$ varies linearly with the applied field, as shown in the inset of Figs.~\ref{fig2}(a) and (b), and is equivalent to Eq.~\ref{eq2} in this limit.

Figures~\ref{fig3}(a-c) report the isotropic and angle-dependent SOT amplitudes as a function of $t_{Pd}$, normalized to $j_{Pd}=10^{7}$~A/cm$^{2}$. To obtain the isotropic FL SOT component, $B^{FL}_{0}$, the Oersted field is calculated as $B^{Oe} = \mu_0 j_{Pd} t_{Pd}/2$ [open triangles in Fig.~\ref{fig3}(a)] and subtracted from the total FL effective field derived from $R_H^{2\omega}$ (open dots). The coefficients $B^{DL\,(FL)}_{0}$, shown in (a), represent the magnetic field induced by the DL (FL) torque when $\mathbf{m} \parallel \mathbf{z}$, whereas $B^{DL\,(FL)}_{2+4}$, shown in (b), represent the angle-dependent contributions, and $B^{DL\,(FL)}_{0+2+4}$, shown in (c), represent the total amplitude of the field when $\mathbf{m} \parallel \mathbf{x} \,(\mathbf{y})$. For comparison with literature data, we plot also the SOT efficiency (right scale), defined by
\begin{equation}
\label{eqSOT_j}
\xi_{j}^{DL\,(FL)} = \frac{2e}{\hbar} M_s t_{Co}\frac{B^{DL\,(FL)}}{j_{Pd}},
\end{equation}
which represents the ratio of the spin current absorbed by the FM to the charge current injected in the NM layer~\cite{LiuPRL2011,NguyenPRL2016}. Remarkably, we find that the ratio $B^{DL}_{0}/j_{Pd}$ increases with increasing $t_{Pd}$, whereas $B^{FL}_{0}/j_{Pd}$ has a non monotonic dependence on $t_{Pd}$. This is a first indication that the interface and bulk contributions to the two torques differ in magnitude. Moreover, we observe that the SOT efficiency depends on the orientation of the magnetization relative to the current direction. In particular, the angular dependence of the FL SOT is much stronger than that of the DL SOT [Fig.~\ref{fig3}(b)], such that the total FL torque is larger than the total DL torque when the magnetization is tilted in-plane [Fig.~\ref{fig3}(c)].

Next, we analyze the SOT amplitudes normalized by the applied electric field $B^{DL,FL}/E$, shown in Fig.~\ref{fig4}(a-c). The motivation for this analysis is that the resistivity of Pd is strongly thickness dependent, as shown in Fig.~\ref{fig1}(b). Therefore, even within a single sample, the current is not homogeneously distributed across the Pd layer. Moreover, the current profile in Pd will vary from sample to sample in a way that is not described by a simple parallel resistor model. The electric field $E=\rho j = \rho_{Pd}j_{Pd}$, on the other hand, is the primary force driving the charge and spin currents~\cite{Freimuth2014PRB} and is constant throughout the thickness of the whole sample, which makes it a practical unit for thickness dependent studies of SOT~\cite{KimNM2013,NguyenPRL2016}.
Accordingly, in analogy with Eq.~\ref{eqSOT_j} and for the purpose of comparing the SOT in different systems, we define the SOT efficiency per unit electric field~\cite{NguyenPRL2016}
\begin{equation}
\label{eqSOT_E}
\xi_{E}^{DL\,(FL)} = \frac{2e}{\hbar} M_s t_{Co}\frac{B^{DL\,(FL)}}{E} = \frac{\xi_{j}^{DL\,(FL)}}{\rho_{Pd}}.
\end{equation}
Figure~\ref{fig4}(a) shows that the thickness dependence of the electric field-normalized SOT is very different from that of the current-normalized SOT: $B^{DL}_0/E$ increases in an almost linear way in the whole range of $t_{Pd}$ while $B^{FL}_{0}/E$ has a monotonic dependence on $t_{Pd}$ and extrapolates to a nonzero value at $t_{Pd}=0$. The angular amplitudes $B^{DL}_{2+4}/E$, on the other hand, are very small relative to $B^{DL}_0/E$ and only weakly thickness dependent, contrary to the FL components $B^{FL}_{2+4}/E$, which have a strong dependence on $t_{Pd}$, as shown in Fig.~\ref{fig4}(b). The implications of these results are discussed below.

\subsection{Discussion of the spin-orbit torque dependence on Pd thickness}
\begin{figure*}
\includegraphics[width = 16cm]{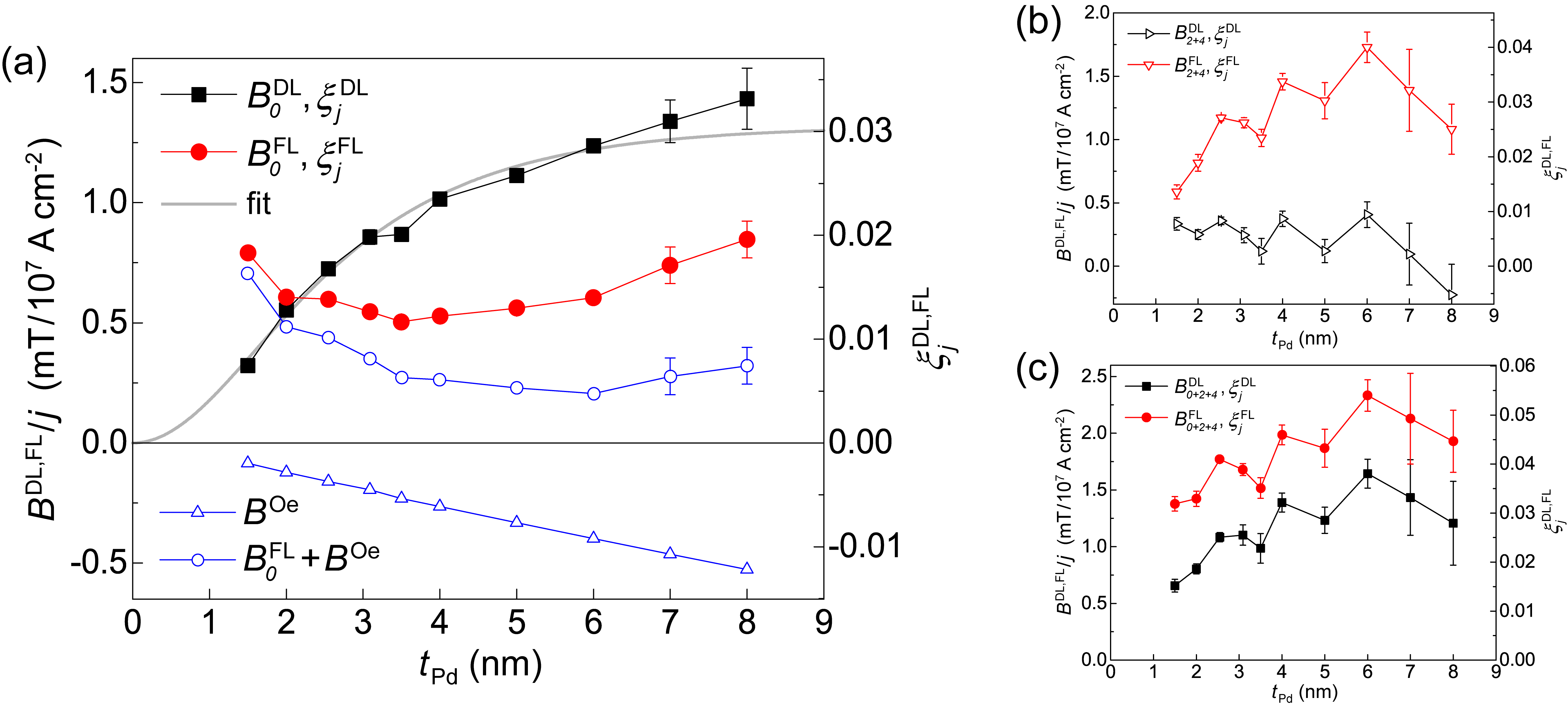}
\caption{SOT fields per unit current density as a function of $t_{Pd}$. (a) Isotropic amplitudes $B^{DL}_{0}/j_{Pd}$ and $B^{FL}_{0}/j_{Pd}$, corresponding to perpendicular magnetization. (b) Angle-dependent amplitudes $B^{DL}_{2+4}/j_{Pd}$ and $B^{FL}_{2+4}/j_{Pd}$. (c) Total amplitudes for in-plane magnetization, $B^{DL}_{0+2+4}/j_{Pd}$ and $B^{FL}_{0+2+4}/j_{Pd}$. The SOT efficiency $\xi_{j}^{DL\,(FL)}$ is shown on the right scale.}\label{fig3}
\end{figure*}

We discuss first the thickness dependence of the isotropic DL torque $B^{DL}_{0}/j_{Pd}$ in terms of the drift-diffusion approach widely employed in the analysis of SOT and spin pumping experiments.  Assuming that the spin current flowing from the NM into the FM is uniquely due to the bulk SHE of the NM and entirely absorbed at the NM/FM boundary~\cite{LiuPRL2011}, the simplest drift-diffusion model gives a dependence of the type $\xi_{j}^{DL} =\gamma_{SH}[1-\mathrm{sech}(t_{Pd}/\lambda)]$. A fit according to this function [grey line in Fig.~\ref{fig3}(a)] yields $\gamma_{SH}=0.03$ and $\lambda=2$~nm, with an incertitude of about 10\%. There are, however, several reasons that caution against drawing conclusions from such a simplified model. First, within drift-diffusion theory, one should take into account the spin backflow into the NM~\cite{HaneyPRB2013,ChenPRB2013} and spin memory loss at the NM/FM interface~\cite{BassAPL2002,rojasPRL2014}. Second, as pointed out in Ref.~\onlinecite{NguyenPRL2016}, drift-diffusion models assume constant $\rho_{NM}$, $\gamma_{SH}$, and $\lambda$ throughout the NM layer, whereas the strong dependence of $\rho_{NM}$ on $t_{NM}$ typical of ultrathin films [Fig.~\ref{fig1}(b)] invalidates this hypothesis. Third, both first-principle electronic calculations~\cite{FreimuthPRB2015a,KellyPRL2016} and optical measurements~\cite{CiccacciPRB2015} suggest the existence of an "interface SHE", which can be significantly larger than the bulk SHE. This interface SHE is ascribed to the current carried by interface states, similarly to the Rashba-Edelstein effect~\cite{SanchezNC2013}. Additionally, spin-dependent scattering at the FM/NM interface can also lead to a spin polarized current and generate DL and FL SOT~\cite{stilesPRB16a,stilesPRB16b}. Keeping track of all these effects together leads to an over-parameterized problem, which makes it difficult to draw a firm conclusion on either $\gamma_{SH}$ or $\lambda$.

An alternative approach is to consider the SOT values normalized by the applied electric field. According to drift-diffusion theory and including spin back flow~\cite{HaneyPRB2013}, the SOT efficiency per unit electric field due to the bulk SHE is given by
\begin{multline}\label{DLSOT}
\xi_{E}^{DL}= 4\lambda\frac{\gamma_{SH}}{\rho} \sinh^{2}\left(\frac{t}{2\lambda}\right)\\
\frac{2\lambda G_i^2\cosh (\frac{t}{\lambda}) + G_r \left[2\lambda G_r \cosh (\frac{t}{\lambda}) +\frac{1}{\rho} \sinh (\frac{t}{\lambda}) \right]}{\left[2G_i \lambda \cosh (\frac{t}{\lambda})\right]^2 + \left[2\lambda G_r \cosh (\frac{t}{\lambda}) +\frac{1}{\rho} \sinh (\frac{t}{\lambda}) \right]^2},
\end{multline}
\begin{multline}\label{FLSOT}
\xi_{E}^{FL}= 4\lambda\frac{\gamma_{SH}}{\rho} \sinh^{2}\left(\frac{t}{2\lambda}\right)\\
\frac{\frac{1}{\rho} G_i \sinh (\frac{t}{\lambda})}{\left[2G_i \lambda \cosh (\frac{t}{\lambda})\right]^2 + \left[2\lambda G_r \cosh (\frac{t}{\lambda}) +\frac{1}{\rho} \sinh (\frac{t}{\lambda}) \right]^2},
\end{multline}
where $\rho$ and $t$ are the resistivity and thickness of the normal metal layer (Pd in our case), while $G_r$ and $G_i$ are the real and imaginary parts of the spin mixing conductance. The grey line in Fig.~\ref{fig3}(c) is a fit of $B^{DL}_{0}/E$ according to Eq.~\ref{DLSOT}, which gives a spin Hall conductivity $\sigma_{SH}=\gamma_{SH}/\rho=(4 \pm 1)\times 10^5$~$\Omega^{-1}$m$^{-1}$ and $\lambda=(7\pm 1)$~nm, assuming $\rho_{Pd}^{\infty} = 1.4 \times 10^9$~$\Omega$m for the bulk Pd resistivity (measured for a 30~nm thick Pd film), and the spin mixing conductance calculated for a permalloy/Pd interface, $G_r = 7.75\times 10^{14}$~$\Omega^{-1}$m$^{-2}$ and $G_i=G_r/7$~\cite{FreimuthPRB2015b}.
Although $\lambda=7$~nm is in good agreement with recent first principles calculations of Pd~\cite{KellyPRB2015}, $\sigma_{SH}$ exceeds the intrinsic value of $2.1\times 10^5$~$\Omega^{-1}$m$^{-1}$ calculated for bulk Pd~\cite{FreimuthPRB2015b} and $\gamma_{SH} = \rho_{Pd}^{\infty} \sigma_{SH} \approx 0.055$ is consistently larger than previously reported for Pd/permalloy bilayers in spin pumping experiments~\cite{Otani-PRB-2011,KondouAPEX2012,tang12apex,Vlaminck2013prb,Silva2013JAP,BassAPL2002,FreimuthPRB2015b}.
An even larger estimate of $\sigma_{SH}$ is obtained if we use $G_r$ to $5.94\times 10^{14}$ calculated for a Co/Pt interface~\cite{HaneyPRB2013}. This analysis suggests that extrinsic or interface-related effects may enhance the DL SOT in Pd/Co/AlO$_x$ relative to what is expected from the bare SHE of Pd. The amplitude of the DL SOT, however, remains smaller than reported for Pt/Co/AlO$_x$~\cite{GarelloNN2013} and Pt/Co/MgO~\cite{NguyenPRL2016}. We remark that the observed variation of the torques cannot be related to changes of the film roughness~\cite{zhouPRB2015}, which is approximately constant across the whole Pd/Co/AlO$_x$ series.
We also note that a refinement of the drift-diffusion model that considers a thickness-dependent $\lambda$ rescaled such that the product $\rho_{Pd}\lambda$ is constant and equal to the bulk value $\rho_{Pd}^{\infty}\lambda^{\infty}$, consistently with the Elliott-Yafet mechanism of spin relaxation~\cite{NguyenPRL2016}, gives $\sigma_{SH}=(5 \pm 2)\times 10^5$~$\Omega^{-1}$m$^{-1}$ and $\lambda=(18\pm 8)$~nm, showing that the fitted values of $\sigma_{SH}$ and $\lambda$ are strongly model dependent. Even such a refined model, however, does not take into account the interface enhancement of the SHE~\cite{FreimuthPRB2015a,KellyPRL2016,stilesPRB16a,stilesPRB16b} and cannot justify the presence of a large FL SOT.

A similar analysis can be performed on $B^{DL}_{0+2+4}$, corresponding to the SOT efficiency for the in-plane magnetization [Fig.~\ref{fig3}(c)]; the main result in this case is a reduction of $\lambda$ by 20-40\% relative to the out-of-plane case discussed above. Such a result is not so surprising if one considers that spin diffusion in the limit of $t_{Pd}\geq\lambda$ is strongly influenced by the NM/FM interface, where spin-orbit coupling is responsible for inducing anisotropic electron scattering processes. For the same reason, one may expect also the spin mixing conductance and spin memory loss to depend on the magnetization direction.

\begin{figure*}
\includegraphics[width = 16cm]{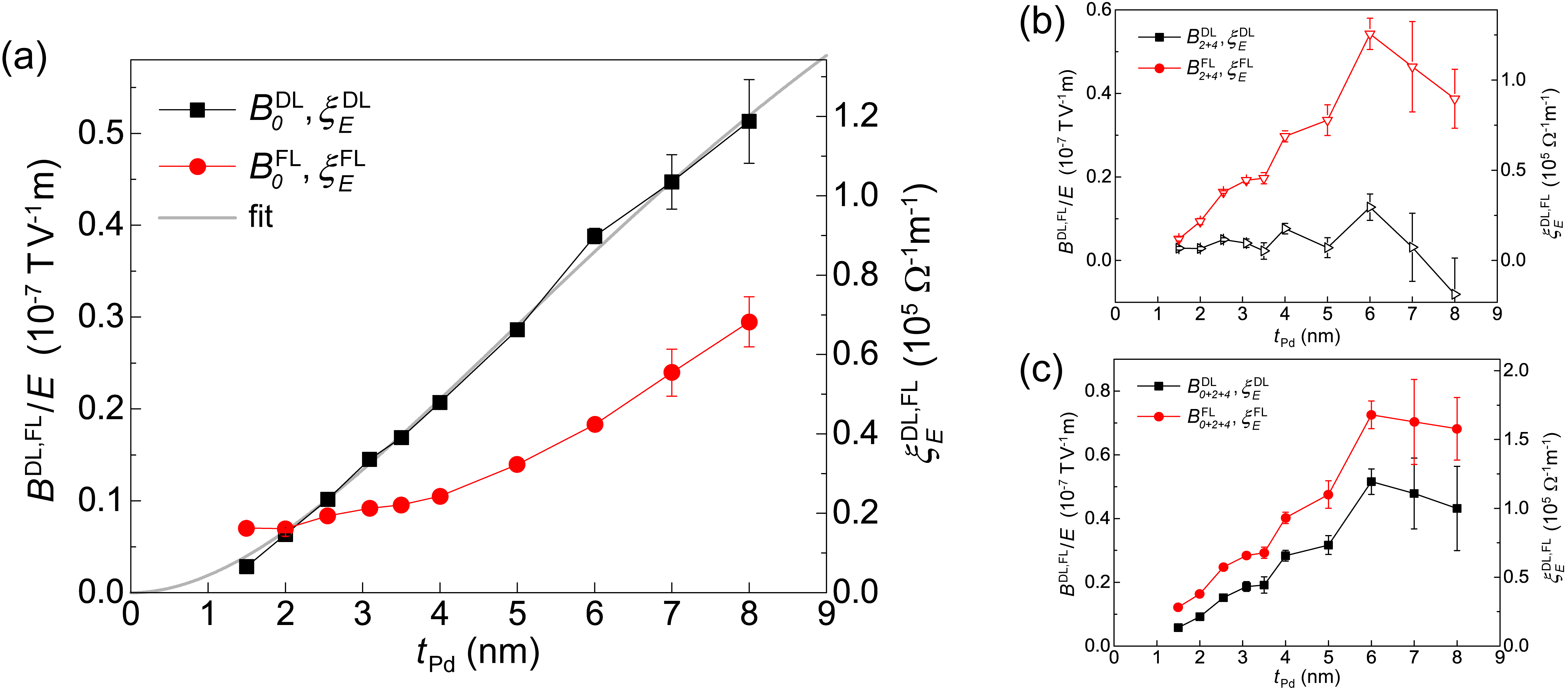}
\caption{SOT fields per unit electric field as a function of $t_{Pd}$. (a) Isotropic amplitudes $B^{DL}_{0}/E$ and $B^{FL}_{0}/E$, corresponding to perpendicular magnetization. (b) Angle-dependent amplitudes $B^{DL}_{2+4}/E$ and $B^{FL}_{2+4}/E$. (c) Total amplitudes for in-plane magnetization, $B^{DL}_{0+2+4}/E$ and $B^{FL}_{0+2+4}/E$. The SOT efficiency $\xi_{E}^{DL\,(FL)}$ is shown on the right scale.}\label{fig4}
\end{figure*}

We now discuss the thickness dependence of the FL SOT, which is qualitatively different compared to the DL SOT. The $B^{FL}_{0}/j_{Pd}$ ratio [full dots in Fig.~\ref{fig3}(a)] has a nonmonotonic behavior as a function of $t_{Pd}$, with a minimum around $3.5$~nm. This behavior compounds the FL SOT thickness dependence with the uneven current distribution in the Pd/Co bilayer. Analysing the ratio $B^{FL}_{0}/E$ [full dots in Fig.~\ref{fig4}(a)] obviates the problem of the current distribution, revealing a monotonic increase of $B^{FL}_{0}/E$ with increasing $t_{Pd}$. Differently from $B^{DL}_{0}/E$, however, $B^{FL}_{0}/E$ extrapolates to a nonzero value at $t_{Pd}=0$, evidencing a significant contribution from the Pd/Co interface akin to a Rashba-Edelstein magnetic field~\cite{ManchonPRB2008,MironNM2010}. Moreover, an attempt to fit $\xi_{E}^{FL}$ using the SHE drift-diffusion theory (Eq.~\ref{FLSOT}) adding a constant term for the interface contribution yields $\sigma_{SH}$ about one order of magnitude larger than derived from the analysis of $\xi_{E}^{DL}$. Similar considerations apply to $B^{FL}_{0+2+4}$. The comparison between $B^{FL}_{2+4}$ and $B^{DL}_{2+4}$ in Fig.~\ref{fig4}(b) further shows that the FL torque has a stronger anisotropic component relative to the DL torque, and that such anisotropy is thickness dependent. This is in contrast with the standard form of the SOT derived from the bulk SHE using either the drift-diffusion model or the Boltzmann transport equation, exemplified by Eqs.~\ref{DLSOT} and \ref{FLSOT}, which only predict the existence of the isotropic terms $B^{DL}_{0}$ and $B^{DL}_{0}$~\cite{HaneyPRB2013}. Although there is presently no theory providing a complete description of the torque anisotropy in realistic FM/HM systems including both bulk and interfacial spin-orbit coupling, a possible explanation is that the angular dependence of the torques arises from anisotropic spin-dependent scattering at the FM/HM interface or inside the FM, which affects the amplitude of the nonequilibrium spin currents in the bilayer.

The observation of a thickness-independent FL SOT and of strongly enhanced DL and FL amplitudes relative to those expected from the bulk SHE of Pd leads to the conclusion that interface effects contribute significantly to the current-induced SOT in Pd/Co/AlO$_x$. Such effects are ubiquitous, but may be particularly evident in Pd because of the reduced bulk SHE compared to 5$d$ metal systems. The standard drift-diffusion theory of the SHE~\cite{HaneyPRB2013,ChenPRB2013}, which is routinely used to extract $\gamma_{SH}$ and $\lambda$ from spin pumping and SOT measurements, does not account for either interface terms, thickness-dependent spin diffusion parameters, or angle-dependent SOT. Similarly, two-dimensional models of the Rashba-Edelstein effect do not properly describe the spin accumulation profile in FM/HM layers and neglect the bulk SHE of the HM~\cite{ManchonPRB2008}. More recent theoretical work consequently points out the need to include the bulk- and interface-generated spin accumulation on the same footing in three-dimensional models of electron transport~\cite{stilesPRB16a,stilesPRB16b}. A key prediction of this later work is that interfacial spin-orbit scattering 
creates substantial spin currents that flow away from the FM/HM interface, generating both DL and FL torques. Interestingly, the thickness dependence of the two torques differs, with the interfacial FL torque being nearly thickness independent and the interfacial DL torque increasing with the thickness of the NM layer~\cite{stilesPRB16b}. This is qualitatively consistent with the results reported in Fig.~\ref{fig4}, namely a FL torque comprising a thickness-independent term and a DL torque that is strongly thickness dependent. In this scenario, both torques would include bulk and interface contributions of comparable magnitude. Another prediction of this theory is that the interfacial scattering amplitudes depend on the magnetization direction, which naturally leads to anisotropic SOT as well as magnetoresistance~\cite{stilesPRB16a,stilesPRB16b}. In this respect it is interesting to draw a parallel between the angular dependence of the SOT and magnetoresistance in metallic FM/HM bilayers. Figures~\ref{fig3} and~\ref{fig4} show that the SOT amplitude increases strongly when $\mathbf{m}$ is tilted towards the $y$ direction and only moderately when $\mathbf{m}$ is tilted towards the $x$ direction. This is similar to the behavior of the magnetoresistance in FM/HM layers, which shows much larger changes when $\mathbf{m}$ is rotated in the $yz$ plane relative to rotations in the $xz$ plane~\cite{KobsPRL2011,AvciNP2015}. Exploring this connection goes beyond the scope of this work, but we believe that additional studies of the correlation between SOT and anisotropic magnetoresistance~\cite{Avci2015APL} may help to understand transport at interfaces with spin-orbit coupling.

\subsection{Magnetization switching}
\begin{figure*}
\includegraphics[width=12cm]{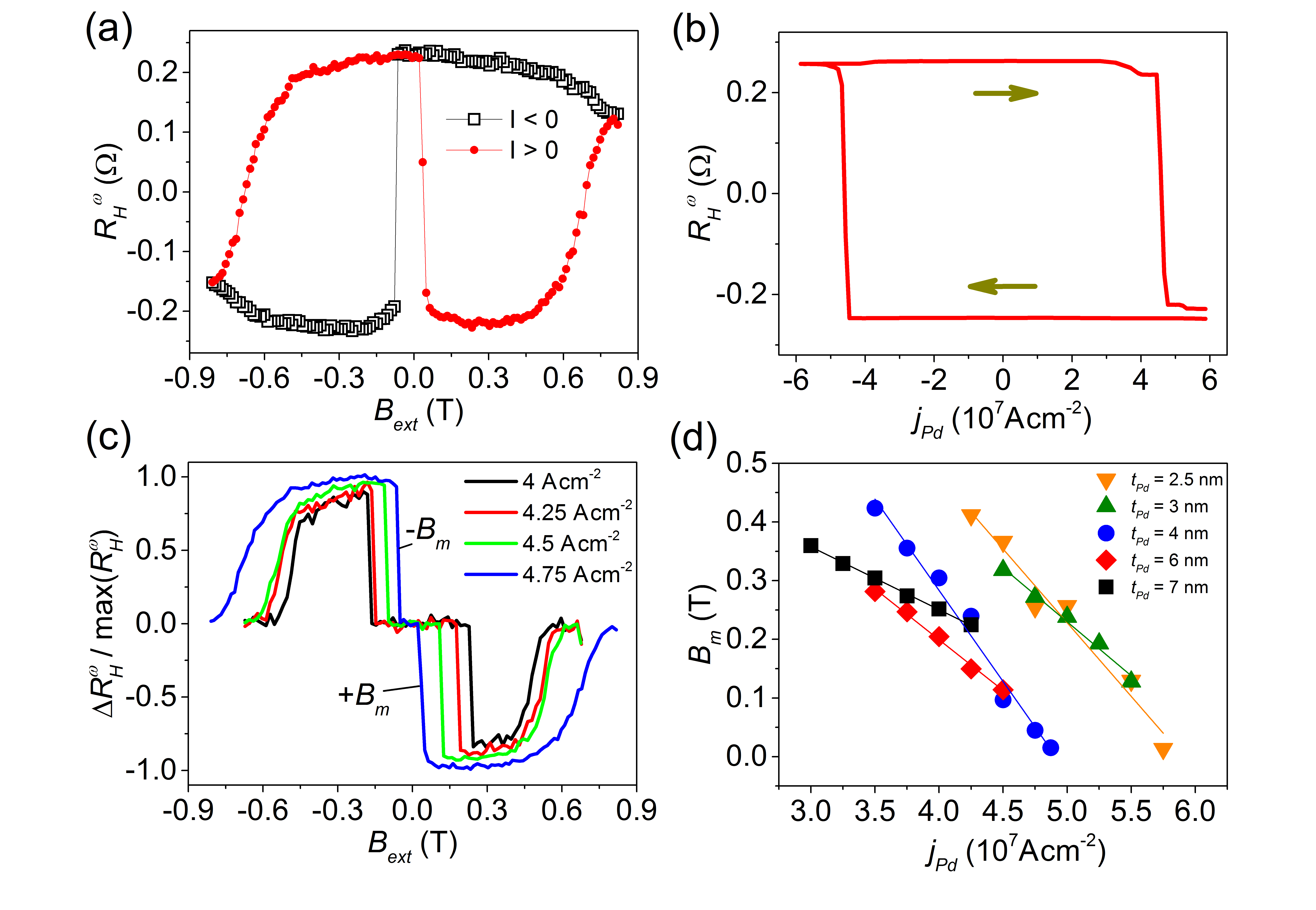}
\caption{(a) Change of the Hall resistance due to magnetization switching induced by positive and negative current
pulses of amplitude $j_{Pd} = 4.75$~A/cm$^2$ ($I= 11$~mA) as a function of external field $B_{ext}$ in Pd(4nm)/Co(0.6nm)/AlO$_x$. The external field is applied in-plane parallel to the current direction ($\theta_B=89.7^{\circ}, \varphi = 0^{\circ}$). (b) Hall resistance as a function of $j_{Pd}$ at constant $B_{ext}=0.2$~T. (c) Normalized difference of the Hall resistance measured for $I>0$ and $I<0$ in (a) as a function of $B_{ext}$. Similar curves are shown also for different values of $j_{Pd}$. (d) Minimum switching field $B_m$ as a function of $j_{Pd}$ and $t_{Pd}$.}\label{fig5}
\end{figure*}

Finally, we address the question of whether the SOT provided by Pd are sufficient to reverse the magnetization of the Co layer in a controlled way. To our knowledge, SOT-driven switching using $4d$ metal layers has not been reported so far. Figure~\ref{fig5}(a) shows the results of a typical switching experiment, performed by injecting 0.3~s long current pulses of positive and negative polarity in a Pd(4nm)/Co(0.6nm)/AlO$_x$ Hall bar. The plot shows the change of the Hall resistance after each pulse due to switching of the magnetization from up to down and vice versa during a single sweep of the in-plane external field, from -0.82 to 0.82~T. An alternative demonstration of current-induced switching is reported in Fig.~\ref{fig5}(b), where the Hall resistance is recorded as a function of current for a constant in-plane field. As discussed in previous work~\cite{MironN2011,AvciAPL2012}, the purpose of the in-plane field is to break the symmetry of the DL SOT and univocally determine the switching direction, which occurs through the expansion of chiral domain walls~\cite{Beach2014APL}. 
Figure~\ref{fig5}(c) shows the difference of the Hall resistance $\Delta R^{\omega}_{H}$ measured after two consecutive current pulses, one positive and one negative, normalised to its saturation value. Switching is obtained whenever $|\Delta R^{\omega}_{H}|>0$. We observe that the minimum field required to achieve deterministic switching, $B_m$, decreases linearly with increasing current, as shown in Fig.~\ref{fig5}(d), similarly to Pt/Co/AlO$_x$~\cite{MironN2011} and Ta/CoFeB/MgO~\cite{AvciPRB2014}. Such a linear behavior is found in all samples, albeit with a different slope, which we attribute to differences in the domain nucleation field. The minimum current density required to achieve deterministic switching depends on $t_{Pd}$, and is generally smaller in the thicker samples, as expected due to the overall increase of the SOT with increasing $t_{Pd}$
shown in Fig.\ref{fig3}. Remarkably, the dc current density required to switch Pd/Co/AlO$_x$ is similar (within a factor two) to that used to switch Pt/Co/MgO and Pt/Co/AlO$_x$ dots with comparable PMA~\cite{AvciAPL2012,GarelloAPL2014}, which is in agreement with the relatively large SOT efficiency reported in this work.

\section{CONCLUSIONS}
In summary, we have performed magnetotransport, magnetic anisotropy, and SOT vector measurements of Pd/Co(0.6nm)/AlO$_x$ layers as a function of Pd thickness. We found that the PMA of Pd/Co is reinforced by optimum oxidation of the Al capping layer relative to Pd/Co multilayers, yielding a magnetic anisotropy energy of $(1.5 \pm 0.1) \times 10^6$~J/m$^3$, which is nearly independent of $t_{Pd}$. Current injection in Pd/Co/AlO$_x$ leads to sizeable DL and FL SOT that have the same order of magnitude, about 1~mT per $10^7$~A/cm$^2$, but different thickness and angular dependence. The analysis of the DL SOT yields a relatively large effective spin Hall angle for Pd, $\gamma_{SH} \approx 0.03-0.06$, or a spin Hall conductivity $\sigma_{SH}=(4 \pm 1)\times 10^5$~$\Omega^{-1}$m$^{-1}$, depending on the model used to fit the data. The DL spin torque efficiency per unit electric field is of the order of $10^5$~$\Omega^{-1}$m$^{-1}$, only a factor of two smaller relative to Pt/Co/MgO layers of comparable thickness~\cite{NguyenPRL2016}. $\gamma_{SH}$ is enhanced compared to Pd/permalloy bilayers~\cite{Otani-PRB-2011,KondouAPEX2012,tang12apex,Vlaminck2013prb,Silva2013JAP}, but significantly smaller than reported for [Pd/Co]$_\mathrm{N}$ multilayers~\cite{Jamali2013PRL} and Pd/FePd/MgO~\cite{Lee2014SR}.
Additionally, our data evidence a strong FL SOT, with an interface contribution that extrapolates to a finite value at $t_{Pd} = 0$, and up to a three-fold enhancement of the FL SOT efficiency when the magnetization rotates from the out-of-plane to the in-plane direction transverse to the current. Overall, our results indicate that SOT models based on one-dimensional drift-diffusion theory and a bulk SHE do not adequately capture the SOT dependence on Pd thickness, at least based on a single set of $\sigma_{SH}$, $\lambda$, $G_r$, and $G_i$ parameters. A possible scenario is that spin currents driven by interfacial spin-orbit scattering add up to the spin currents induced by the bulk SHE of Pd~\cite{stilesPRB16b}, resulting in the nontrivial thickness and angle dependence of the SOT observed here.
Finally, we report bipolar magnetization switching in Pd/Co/AlO$_x$ for $j_{Pd}= 3 - 6\times 10^7$~A/cm$^2$, depending on the Pd thickness and in-plane applied field. These results show that Pd/Co/oxide layers with relatively low resistivity can be used to combine strong PMA with efficient current-induced magnetization switching, opening the possibility of using Pd as an alternative material to Pt, Ta, and W in SOT devices.


\section{ACKNOWLEDGMENTS}
This work was supported by the Swiss National Science Foundation (Grant No. 200021-153404) and the European Commission under the Seventh Framework Program (spOt project, Grant No. 318144). We thank Junxiao Feng and Luca Persichetti for performing the roughness measurements.


%

%

\end{document}